\documentclass[longauth]{aa}  

\usepackage{graphicx}
\usepackage{txfonts}
\usepackage{siunitx}						
\usepackage{physics}
\usepackage{hyperref}
\usepackage{multicol}
\usepackage{lipsum}

\usepackage{xcolor}

\newcommand\solarmass{M_\odot}

\hypersetup{
	colorlinks			= true			, % Set to false to disable coloring links
	citecolor			= blue!80!black			, % The color of citations
	linkcolor			= blue			, % The color of references to document (sections ,figures, etc)
	urlcolor			=blue!80!black		, % The color of hyperlinks (URLs)
	breaklinks			=true           ,
}

\usepackage{natbib}
\bibpunct{(}{)}{;}{a}{}{,}             
\begin{document}

\title{Polarimetry and astrometry of NIR flares as event horizon scale, dynamical probes for the mass of Sgr~A*}
 	
 	\author{
  	    The GRAVITY Collaboration\fnmsep\thanks
  	    {
  	    GRAVITY is developed in a collaboration by MPE, LESIA of Paris Observatory / CNRS / Sorbonne Université / Univ.
  	    Paris Diderot and IPAG of Université Grenoble Alpes / CNRS, MPIA, Univ. of Cologne, CENTRA - Centro de
  	    Astrofisica e Gravitação, and ESO. Corresponding authors:
  	    A.~Drescher (drescher@mpe.mpg.de), D.~C.~Ribeiro (dribeiro@mpe.mpg.de) \& N.~Aimar (nicolas.aimar@obspm.fr). 
  	    }:
  	    R.~Abuter                   \inst{4}            \and
  	    N.~Aimar                    \inst{2}            \and
  	    P.~Amaro Seoane             \inst{18,1,25,19}   \and
  	    A.~Amorim                   \inst{8,7}          \and
  	    M.~Bauböck                  \inst{16}           \and
  	    J.P.~Berger                 \inst{3,4}          \and
  	    H.~Bonnet                   \inst{4}            \and
  	    G.~Bourdarot                \inst{1}            \and
  	    W.~Brandner                 \inst{5}            \and
  	    V.~Cardoso                  \inst{7,21}         \and
  	    Y.~Clénet                   \inst{2}            \and
  	    R.~Davies                   \inst{1}            \and
  	    P.T.~de~Zeeuw               \inst{24}           \and
  	    J.~Dexter                   \inst{11}           \and
  	    A.~Drescher                 \inst{1}            \and
  	    A.~Eckart                   \inst{6,17}         \and
  	    F.~Eisenhauer               \inst{1,23}         \and
  	    H.~Feuchtgruber             \inst{1}            \and
  	    G.~Finger                   \inst{1}            \and
  	    N.M.~Förster~Schreiber      \inst{1}            \and
  	    A.~Foschi                   \inst{7,10}         \and
  	    P.~Garcia                   \inst{12,7}         \and
  	    F.~Gao                      \inst{13,1}         \and
  	    Z.~Gelles                   \inst{26}           \and
  	    E.~Gendron                  \inst{2}            \and
  	    R.~Genzel                   \inst{1,14}         \and
  	    S.~Gillessen                \inst{1}            \and
  	    M.~Hartl                    \inst{1}            \and
  	    X.~Haubois                  \inst{9}            \and
  	    F.~Haussmann                \inst{1}            \and
  	    G.~Heißel                   \inst{20,2}         \and
  	    T.~Henning                  \inst{5}            \and
  	    S.~Hippler                  \inst{5}            \and
  	    M.~Horrobin                 \inst{6}            \and
  	    L.~Jochum                   \inst{9}            \and
  	    L.~Jocou                    \inst{3}            \and
  	    A.~Kaufer                   \inst{9}            \and
  	    P.~Kervella                 \inst{2}            \and
  	    S.~Lacour                   \inst{2}            \and
  	    V.~Lapeyrère                \inst{2}            \and
  	    J.-B.~Le~Bouquin            \inst{3}            \and
  	    P.~Léna                     \inst{2}            \and
  	    D.~Lutz                     \inst{1}            \and
  	    F.~Mang                     \inst{1}            \and
  	    N.~More                     \inst{1}            \and
  	    T.~Ott                      \inst{1}            \and
  	    T.~Paumard                  \inst{2}            \and
  	    K.~Perraut                  \inst{3}            \and
  	    G.~Perrin                   \inst{2}            \and
  	    O.~Pfuhl                    \inst{4,1}          \and
  	    S.~Rabien                   \inst{1}            \and
  	    D.~C.~Ribeiro                \inst{1}            \and
  	    M.~Sadun Bordoni            \inst{1}            \and
  	    S.~Scheithauer              \inst{5}            \and
  	    J.~Shangguan                \inst{1}            \and
  	    T.~Shimizu                  \inst{1}            \and
  	    J.~Stadler                  \inst{15,1}         \and
  	    O.~Straub                   \inst{1,22}         \and
  	    C.~Straubmeier              \inst{6}            \and
  	    E.~Sturm                    \inst{1}            \and
  	    L.J.~Tacconi                \inst{1}            \and
  	    F.~Vincent                  \inst{2}            \and
  	    S.~von~Fellenberg           \inst{17,1}         \and
  	    F.~Widmann                  \inst{1}            \and
  	    M.~Wielgus                  \inst{17}           \and
  	    E.~Wieprecht                \inst{1}            \and
  	    E.~Wiezorrek                \inst{1}            \and
  	    J.~Woillez                  \inst{4}            
    }

    \institute{
  	    %1 
  	    Max Planck Institute for Extraterrestrial Physics, Giessenbachstraße 1, 85748 Garching, Germany \and
  	    %2
  	    LESIA, Observatoire de Paris, Université PSL, CNRS, Sorbonne Université, Université de Paris, 5 place Jules Janssen, 92195 Meudon, France \and 	
  	    %3
  	    Univ. Grenoble Alpes, CNRS, IPAG, 38000 Grenoble, France \and
  	    %4
  	    European Southern Observatory, Karl-Schwarzschild-Straße 2, 85748 Garching, Germany \and
  	    %5
  	    Max Planck Institute for Astronomy, Königstuhl 17, 69117 Heidelberg, Germany \and
  	    %6
  	    1st Institute of Physics, University of Cologne, Zülpicher Straße 77, 50937 Cologne, Germany \and
  	    %7
  	    CENTRA - Centro de Astrofísica e Gravitação, IST, Universidade de Lisboa, 1049-001 Lisboa, Portugal \and
  	    %8
  	    Universidade de Lisboa - Faculdade de Ciências, Campo Grande, 1749-016 Lisboa, Portugal \and
  	    %9
  	    European Southern Observatory, Casilla 19001, Santiago 19, Chile \and
  	    %10
  	    Universidade do Porto, Faculdade de Engenharia, Rua Dr. Roberto, Frias, 4200-465 Porto, Portugal \and
  	    %11
  	    Department of Astrophysical \& Planetary Sciences, JILA, Duane Physics Bldg., 2000 Colorado Ave, University of Colorado, Boulder, CO 80309, USA 
  	    \and
  	    %12
  	    Faculdade de Engenharia, Universidade do Porto, rua Dr. Roberto Frias, 4200-465 Porto, Portugal \and
  	    %13
  	    Hamburger Sternwarte, Universität Hamburg, Gojenbergsweg 112, 21029 Hamburg, Germany \and
  	    %14
  	    Departments of Physics \& Astronomy, Le Conte Hall, University of California, Berkeley, CA 94720, USA \and
  	    %15
  	    Max Planck Institute for Astrophysics, Karl-Schwarzschild-Straße 1, 85748 Garching, Germany \and
  	    %16
  	    Department of Physics, University of Illinois, 1110 West Green Street, Urbana, IL 61801, USA \and
  	    %17
  	    Max Planck Institute for Radio Astronomy, auf dem Hügel 69, 53121 Bonn, Germany \and
  	    %18
  	    Institute of Multidisciplinary Mathematics, Universitat Politècnica de València, València, Spain \and
  	    %19
  	    Kavli Institute for Astronomy and Astrophysics, Beijing, China \and
  	    %20
  	    Advanced Concepts Team, ESA, TEC-SF, ESTEC, Keplerlaan 1, 2201 AZ Noordwijk, The Netherlands \and
  	    %21
  	    Niels Bohr International Academy, Niels Bohr Institute, Blegdamsvej 17, 2100 Copenhagen, Denmark \and
  	    %22
  	    ORIGINS Excellence Cluster, Boltzmannstraße 2, 85748 Garching, Germany \and
  	    %23
  	    Department of Physics, Technical University of Munich, 85748 Garching, Germany \and
  	    %24
  	    Leiden University, 2311EZ Leiden, The Netherlands \and
  	    %25
  	    Higgs Centre for Theoretical Physics, Edinburgh, UK \and
  	    %26
  	    Department of Physics, Jadwin Hall, Washington Road, Princeton, New Jersey 08544, USA
    }  

    \date{Aug 31, 2023}
  
    \abstract{
    We present new astrometric and polarimetric observations of flares from Sgr~A* obtained with GRAVITY, the near-infrared interferometer at ESO's Very Large Telescope Interferometer (VLTI), bringing the total sample of well-covered astrometric flares to four and polarimetric flares to six. Of all flares, two are well covered in both domains. All astrometric flares show clockwise motion in the plane of the sky with a period of around an hour, and the polarization vector rotates by one full loop in the same time. Given the apparent similarities of the flares, we present a common fit, taking into account the absence of strong Doppler boosting peaks in the light curves and the EHT-measured geometry. Our results are consistent with and significantly strengthen our model from 2018. First, we find that the combination of polarization period and measured flare radius of around nine gravitational radii ($9 R_g \approx 1.5 R_{ISCO}$, innermost stable circular orbit) is consistent with Keplerian orbital motion of hot spots in the innermost accretion zone. The mass inside the flares' radius is consistent with the \SI{4.297e6}{\solarmass} measured from stellar orbits at several thousand $R_g$. This finding and the diameter of the millimeter shadow of Sgr~A* thus support a single black hole model. Second, the magnetic field configuration is predominantly poloidal (vertical), and the flares' orbital plane has a moderate inclination with respect to the plane of the sky, as shown by the non-detection of Doppler-boosting and the fact that we observe one polarization loop per astrometric loop. Finally, both the position angle on the sky and the required magnetic field strength suggest that the accretion flow is fueled and controlled by the winds of the massive young stars of the clockwise stellar disk 1-5 $\arcsec$ from Sgr~A*, in agreement with recent simulations.
    }
  
  
    \titlerunning{Dynamical probes for the mass of Sgr~A*}
    \authorrunning{GRAVITY Collaboration}
    \maketitle

%============================================================

\section{Introduction}
\label{sec:introduction}

The presence of a massive black hole (MBH) in the Galactic center (GC) has been established, beyond any reasonable doubt, in a four-decade-long research by several teams 
\citep[e.g., ][]{Genzel_2021}. Not only is Sagittarius A* (Sgr~A*) the best case for the existence of black holes, but unlike any other known black hole, it can be resolved both spatially and temporally, allowing dynamical tests long considered unfeasible. The combination of precise stellar astrometry and spectroscopy has revealed the gravitational redshift signature in the orbit of the star S2 \citep{GRAVITY_2018, Do_2019a}.
Furthermore, the prograde relativistic precession of the orbit yielding an astrometric signature of around 0.5 mas per revolution was detected employing the ultra-high resolution of the GRAVITY interferometer \citep{GRAVITY_2020a,GRAVITY_2022}. Emission from a distance of just a few $R_g$ ($R_g = G M / c^2$ corresponds to \SI{5.12}{\micro as} at our assumed $R_0 = \SI{8.277}{kpc}$ and for $M = \SI{4.297e6}{\solarmass}$ from \citealt{GRAVITY_2022}) was detected in the X-ray \citep{Baganoff_2001,Neilsen_2013}, near-infrared (NIR, \cite{Genzel_2003a,Eisenhauer_2005,Dodds_Eden_2011,Do_2019b,Witzel_2021}), and radio bands \citep{Balick_1974,Lo_1998,Krichbaum_1998,Shen_2005,Yusef_Zadeh_2006,Doeleman_2008,Johnson_2015,EHT_2022}. The small spatial scales imply that the light arises deep in the strong gravitational field of the MBH, and that its propagation is subject to strong relativistic effects.

The NIR emission of Sgr~A* is highly variable, with flares of ten times the quiescent flux occurring once or twice per day \citep{Dodds_Eden_2011,Witzel_2018,GRAVITY_2020b}. The flare emission most likely comes from locally heated electrons emitting synchrotron emission from the NIR, possibly up to the X-ray band \citep{Yuan_2003,Dodds_Eden_2009,Ponti_2017,GRAVITY_2021b}. The synchrotron model with emission coming from a compact source also explains the observed variable NIR polarization of a few tens of percent linear polarization and a negligible amount of circular polarization \citep{Eckart_2006,Eckart_2008,Trippe_2007,Zamaninasab_2010,Witzel_2010,Shahzamanian_2015,GRAVITY_2020a}, as well as the co-evolution of the light curves in the two bands. The estimated source sizes of around 1 $R_g$ and the required magnetic field strengths of around \SI{30}{}-\SI{80}{G} \citep{Dodds_Eden_2010,Eatough_2013,Ripperda_2020} are in agreement with the ``hot spot model'' with orbital radii close to the innermost circular orbit \citep{Broderick_2005,Broderick_2006,Meyer_2006,Hamaus_2009,Ripperda_2020,GRAVITY_2020d}. This model has observable signatures in the astrometry and polarimetry of the NIR emission, where the astrometric amplitudes of a few \SI{10}{\micro as} were one key driver for the development of GRAVITY \citep{Paumard_2006,GRAVITY_2017}. Moreover, \cite{Dexter_2020} show by means of general relativistic magnetohydrodynamic simulations that for magnetically arrested disks (MADs), flares with properties similar to the observed ones might arise from magnetic reconnection events.

GRAVITY combines the light from either the four Auxiliary Telescopes (\SI{1.8}{\meter} diameter) or the four Unit Telescopes (\SI{8}{\meter} diameter) at the Paranal observatory. The spatial resolution in K-band is around \SI{3}{mas}, and the accuracy by which sources can be located is of a few \SI{10}{ \micro as} -- good enough to follow the motions of orbiting hot spots. In 2018 we observed in three nights flares from Sgr~A*, all of which showed astrometrically clockwise motions at radii compatible with the hot spot model. For one flare we also obtained polarimetry in 2018, showing a full rotation of the polarization vector \citep{GRAVITY_2018b}.

Clearly, the number of such high-precision flares was still small. Hence, we embedded in our nightly observing strategy regular visits to Sgr~A* and analyzed in real time the incoming data to monitor for flux increases. If a flare occurred, we continued to observe Sgr~A*. 

Here we report on one new flare, for which the astrometric and polarimetric data are good enough to resolve orbital motion, and on four additional flares, for which only the polarimetry is of sufficient quality.	

%============================================================

\section{Observations}
\label{sec:observations}

We observe the GC typically six months per year, with a total allocation of around 90 hours. The data serve for the astrometric monitoring of the stellar orbits \citep{GRAVITY_2018,GRAVITY_2022} and for observing the emission from Sgr~A* itself \citep{GRAVITY_2018b}. We report on as-yet-unpublished data obtained from observations in 2019, 2021, and 2022 during which Sgr~A* exhibited flares. We analyzed them together with previous flare observations. Specifically, we used the observations listed in Table~\ref{tab:observations}.
\begin{table}
	\caption[]{List of observations with astrometric~(A) and polarimetric~(P) data. The Program IDs for the 2018 data are 0101.B-0576(C) and 0101.B-0576(E). For 2019, 2020, and 2022 they are respectively 0103.B-0032(C), 105.20B2.004, and 109.22ZA.002. }
	\label{tab:observations}
	\centering
	\begin{tabular}{cccclc}
		\hline \hline
		\noalign{\smallskip}
		Date of night & UTC interval & \# of frames & A & P\\
		\noalign{\smallskip}
		\hline
		\noalign{\smallskip}
		27 \makebox[20pt][c]{May} 2018 & 07:02 -- 09:15 & 19 & x & 	 \\
		22 \makebox[20pt][c]{Jul} 2018 & 02:44 -- 04:51 & 19 & x & 	 \\
		28 \makebox[20pt][c]{Jul} 2018 & 00:02 -- 04:16 & 36 & x & x \\
		18 \makebox[20pt][c]{Aug} 2019 & 23:46 -- 03:46 & 32 &   & x \\
		26 \makebox[20pt][c]{Jul} 2021 & 23:25 -- 03:47 & 36 &   & x \\
		27 \makebox[20pt][c]{Jul} 2021 & 23:29 -- 03:21 & 33 &   & x \\
		19 \makebox[20pt][c]{May} 2022 & 04:14 -- 09:41 & 45 & x & x \\
		16 \makebox[20pt][c]{Aug} 2022 & 23:40 -- 02:06 & 26 &   & x \\
		\noalign{\smallskip}
		\hline 
	\end{tabular}
\end{table}
%

%============================================================

\section{Data}
\label{sec:data}

Our GRAVITY data deliver position and polarization measurements simultaneously. The former are obtained from the phase measurements, the latter by comparing (interferometric) fluxes behind a Wollaston prism.

%------------------------------
\subsection{Astrometry}
\label{subsec:data_astrometry}
%------------------------------

For the GC, GRAVITY uses optical fibers to simultaneously observe two objects and record the interference of the two targets on the six baselines formed by the four \SI{8}{\meter} telescopes. One target is the star IRS 16C ($m_K = 9.7$), used as a phase reference, and the other is the science target. Each GRAVITY fiber has a field of view of approximately \SI{70}{mas}, comparable to the diffraction limit of the individual telescopes. 

Since the beginning of the GRAVITY observations, stars were always close enough to Sgr~A* to be observed in the same field of view, yielding binary, triple, or multiple signatures in the interference pattern. In 2018 the star S2 ($m_K = 14.0$) was observed together with Sgr~A* in this way. In this single-beam measurement, information such as the position vector and flux ratio between the two sources, is directly inferred from the measured closure phases and amplitudes. Since 2019, S2 has moved so far out of the interferometric field of view that other fainter stars have been detected close to Sgr~A* \citep{GRAVITY_2021a,GRAVITY_2022}. However, during a flare, Sgr~A* outshines these fainter stars and effectively appears as a single source. 
In this case the astrometry of Sgr~A* has to be derived from the visibility phases. These are measured relative to the reference target via GRAVITY's metrology system, which traces the optical path differences between the telescopes and the instrument~\citep[Appendix A therein]{GRAVITY_2020a}.

Hence, for the new, dual-beam data sets, the astrometry is obtained as a sum of three measured phases: of Sgr~A*, of IRS 16C, and the metrology linkage. The source models are simply unary fits in this case. In practice, we measure the Sgr~A* phases and, every few exposures, that of a local phase reference for which we chose S2. Since
\begin{equation}
	\Big( \varphi_{Sgr~A*} - \varphi_{IRS16C} \Big) 
	- 
	\Big( \varphi_{S2} - \varphi_{IRS16C} \Big)
	=
	\Big( \varphi_{Sgr~A*} - \varphi_{S2} \Big) \; , 
\end{equation}
we reference the Sgr~A* positions to S2. The proper motion of S2 during a flare is small enough not to affect our measurements.

A significant systematic error in our dual-beam flare astrometry is due to high-order imperfections in the GRAVITY optics. As the metrology sensors cover only four points of the pupil, the high-order aberrations translate into phase errors, and thus astrometric uncertainty. These errors average out effectively by the pupil rotation over the course of the night so that stellar orbits are routinely measured with an accuracy down to $\SI{20}{}-\SI{50}{\micro as}$~\cite[e.g.,][]{GRAVITY_2022}.
For the individual short exposures of flare observations, however, these metrology footprint errors dominate over the statistical errors. Since 2022, we circumvent this source of error by simultaneously modulating the telescope pupils relative to the metrology receivers such that the high-order aberrations average out. This restores the astrometry to a level similar to what we achieved in the single-beam case. This new observing mode was not yet in place in 2019 and 2021, and the accuracy of the dual-beam astrometry was not at the level needed for flare orbits during this period.

%------------------------------
\subsection{Polarimetry}
\label{subsec:data_polarimetry}
%------------------------------

In its polarimetric mode, GRAVITY splits the light post-beam combination by means of a Wollaston prism into two orthogonal, linear polarization states. Employing a half-wave plate, we rotate the polarization by \ang{45} between exposures, such that for each pair of exposures, we measure the polarized flux in the \ang{0}, \ang{45}, \ang{90}, and \ang{135} directions. 
The polarization angles are measured with respect to the equatorial system and defined between \ang{0} and \ang{180} east of north.

From the polarized fluxes, the Stokes parameter $Q$ and $U$ are calculated as
\begin{equation}
	Q^\prime = \frac{f_0 - f_{90}}{f_{0} + f_{90}} \quad ,\quad U^\prime = \frac{f_{135} - f_{45}}{f_{135} + f_{45}} \quad ,
\end{equation}
where $f_\theta$ represents the correlated flux along direction $\theta$ and the primes indicate that the values are measured in the detector coordinate system. To get the on-sky polarization state $Q$ and $U$, it is necessary to correct for the geometric beam propagation and the instrumental polarization. This is done with the model developed by \cite{GRAVITY_2023b}. As GRAVITY cannot measure circular polarization, we correct for birefringence with the model presented by \cite{Witzel_2010}, using the fact that the polarized NIR emission of Sgr~A* is predominantly linearly polarized.

Our $Q$ and $U$ values are fractional values normalized to the observed intensity. 
Unlike at longer wavelengths, the variations of the NIR polarized fluxes are completely dominated by the overall brightness variations -- a result of the much shorter heating and cooling timescales in the NIR \citep{von_Fellenberg_2023}. Thus, the fractional values carry the information on the magnetic field and geometry of the Sgr~A* system.

Following \cite{GRAVITY_2020b}, there are two different ways to measure the required fluxes. In the first, if Sgr~A* is the dominant (or only) source in the (interferometric) field of view, the coherent flux directly measures the flux of Sgr~A*. In the second, if there are multiple sources in the field of view, a multiple source fit to these sources includes the flux ratio of Sgr~A* to each star, which, when multiplied with the coherent flux, yields the desired Sgr~A* flux.
Flux measurements for observations up to 2020 are reported in \cite{GRAVITY_2020b}. For 2021 and 2022 we used binary fits with the stars S29 (2021) and S38 (2022) as the second object beyond Sgr~A* (see, e.g., \cite{GRAVITY_2022b} for images of the central region where the stars S29 and S38 are shown).

%============================================================

\section{Analysis}
\label{sec:analysis}

\begin{figure*}
	\centering
	\includegraphics[width=0.99\linewidth]{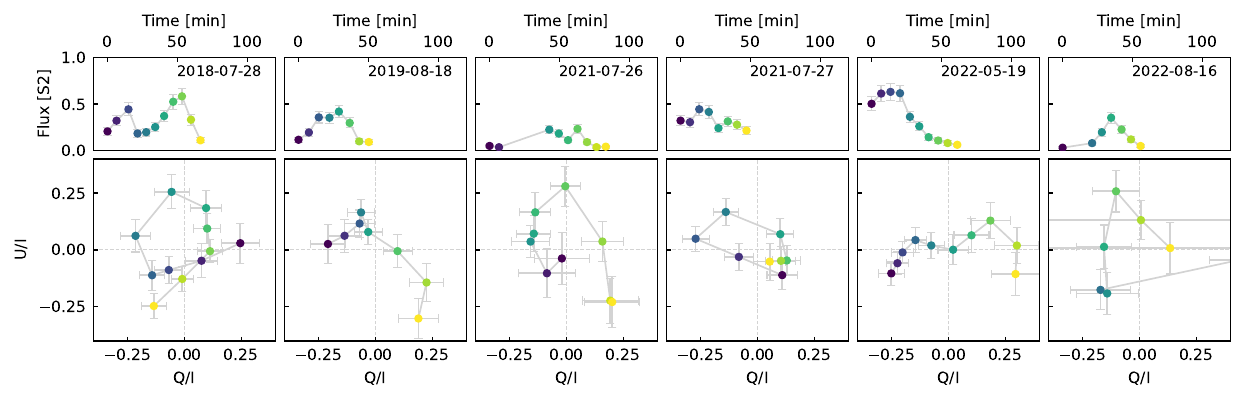}
	\caption{Flares with polarimetric signature. Top row: Flux evolution for each flare. Bottom row: Polarimetric signal in $Q/I$ and $U/I$, color-coded according to time. The flare from 28 July 2018 was published and analyzed in \cite{GRAVITY_2018b,GRAVITY_2020c}. The other five are newly reported here.}
	\label{fig:polarization_flares}%
\end{figure*}

\begin{figure*}
	\centering
	\includegraphics[width=0.7\linewidth]{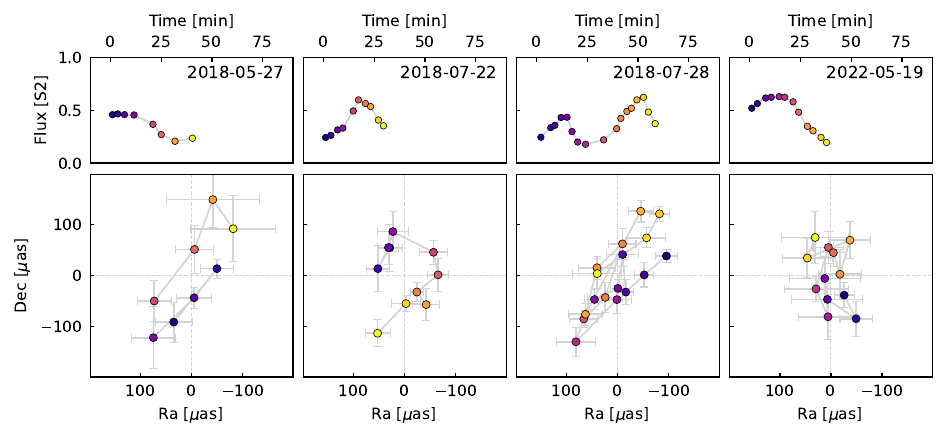}
	\caption{Flux and centroid measurement of the four astrometric flares from 2018 and 2022. The three 2018 flares were published in \cite{GRAVITY_2018b,GRAVITY_2020c}.}
	\label{fig:astrometry_flares}
\end{figure*}

\begin{figure*}
	\centering
	\includegraphics[width=0.947\linewidth]{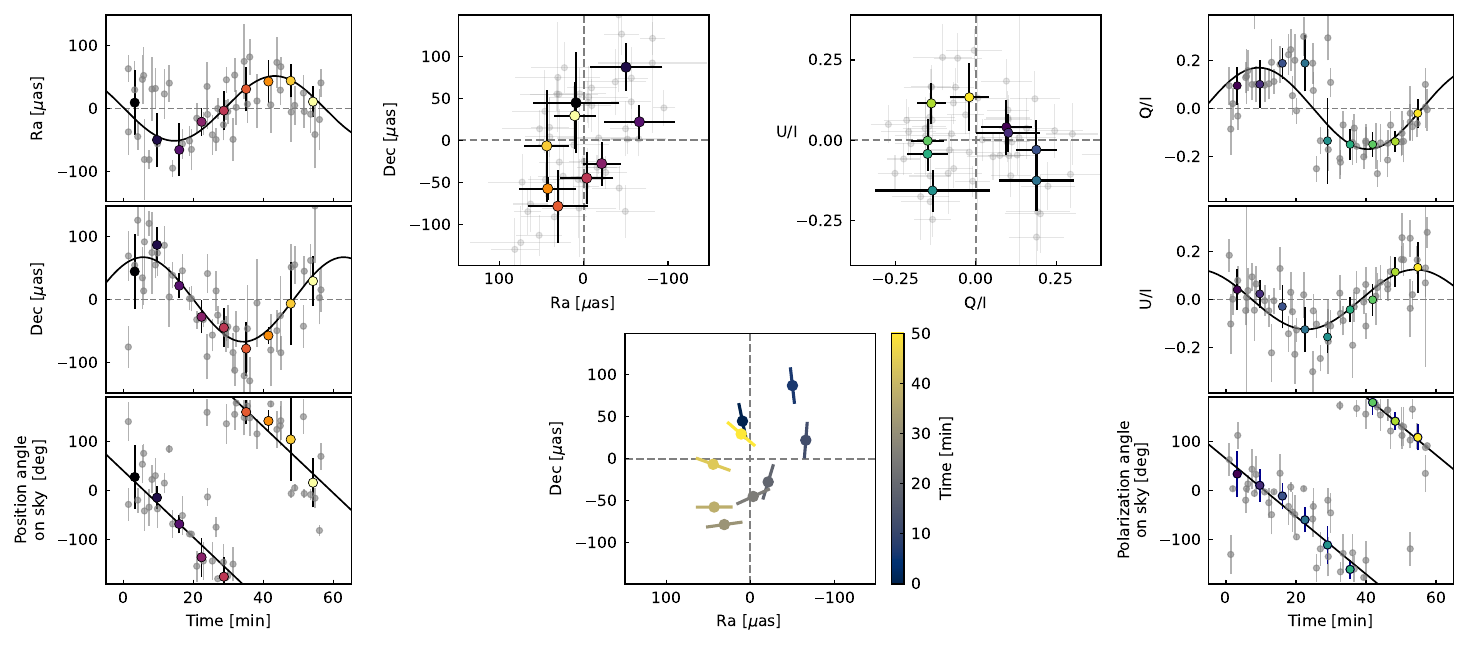}
	\caption{Combined astrometric (left) and polarimetric (right) data. The outer left panels show R.A., Dec., and position angle on the sky as a function of time. The outer right panels show $Q/I$, $U/I$, and polarization angle on the sky vs. time. The full data are shown in gray, the colored points are bins of five minutes, and the color indicates time. The binned data are obtained after wrapping all points around the fit period of $\SI{60}{minutes}$, and the error bars are the standard deviation of the binned data. Overplotted on the angle plots are slopes of $\SI{6}{\degree/\minute} = \SI{360}{\degree/hour}$. The top middle panels illustrate the loops on the sky (left) and in the $Q-U$ plane (right). The bottom middle panel shows the rotation of the polarization for the corresponding astrometric points, one polarization rotation per astrometric orbit. The electric field vector rotates clockwise in the plane of the sky, and in the $Q-U$ plane.
	}
	\label{fig:averaged_flares}%
\end{figure*}

Our sample of flares with polarimetric coverage has grown from one \citep{GRAVITY_2018b,GRAVITY_2020c} to six, all of which show a single clockwise loop, or part thereof, in the $Q-U$ plane~(Fig.~\ref{fig:polarization_flares}). The sample of astrometric measurements has also grown and consists now of four events: the three flares from \cite{GRAVITY_2020b} and the addition from 2022 May 19 (Fig.~\ref{fig:astrometry_flares}).

In contrast to the flare astrometry, where we need a precision of $\approx \SI{10}{}-\SI{30}{\micro as}$ per five-minute exposure, polarimetry is less sensitive to observing conditions and is not penalized by the optical imperfections. As a result, we obtained valid polarimetry for flares even if the conditions were not good enough for astrometry. This is the case for the flare from 2019, the two from 2021, and the one from August 2022.

The similarity between the flares motivated us to look at them in an averaged way. We combined the data of the six polarimetric and four astrometric flares by allowing only a relative time shift between the individual events and finding the solution that best represents a sinusoidal in both sky coordinates for the combined flare data. This yields the data shown in Fig.~\ref{fig:averaged_flares}. Both averaged data sets are very well described by elliptical figures with a common period of around 60 minutes and constant angular velocity of \SI{6}{\degree/\minute}.

The mean astrometric loop size in Fig.~\ref{fig:averaged_flares} of around $r=61\pm9\; \SI{ }{\micro as}$ (corresponding to $0.50 \pm 0.07 \; \SI{}{AU}$ for our assumed $R_0$) and the polarization period of $P = 60 \pm 3 \; \SI{}{\min}$ allows us to estimate the enclosed mass $M_{enc}$. Assuming Keplerian orbital motion, this yields $M_{enc,estimate} = (5.1 \pm 1.6) \times \SI{e6}{\solarmass}$ (Appendix \ref{appendix:mass_estimate}). Within the error, this value agrees with that known from stellar orbits, $M = (4.297 \pm 0.012) \times \SI{e6}{\solarmass}$ \citep{GRAVITY_2022} and shows that (within the uncertainties) the mass of Sgr~A* is enclosed within the flare orbit.
	
	In the following we refine the value of $M_{encl}$ by modeling the astrometry and polarimetry of the flares. We model the astrometric motion of the combined data set as in \cite{GRAVITY_2020d} by an orbiting relativistic hot spot, including the effects of lensing via ray tracing. However, here we enhance the pure astrometric fit by including more information, mainly from the polarimetry.
First, we identify the period of the polarization loop with that of the astrometric loop. Technically this sets a prior on the orbital period. We use $P_{pol} = 60 \pm 3 \; \SI{}{\minute}$. 
Second, we show that observing a single polarization loop per astrometric loop yields a strong prior on the allowed range of inclinations (see Sect.~\ref{subsec:analysis_modeling_the_polarization_loop}). The angular velocity of \SI{6}{\degree/min} constrains the inclination to $i = \SI{157}{\degree} \pm \SI{5}{\degree}$.
Third, the absence of strong spikes from Doppler beaming in the light curves also limits the inclination to be close to face-on. Following \cite{Hamaus_2009} and \cite{GRAVITY_2018} (Appendix C therein), we place a conservative prior of $i = \SI{180}{\degree} \pm \SI{40}{\degree}$.
	Finally, the Event Horizon Telescope (EHT) image of Sgr~A* constrains the inclination to $i = \SI{180}{\degree} \pm \SI{50}{\degree}$ \citep{EHT_2022}.

We took into account all these constraints by including them in the likelihood function and determined a globally best-fitting orbiting hot spot model. The parameters of the fit are the hot spot's orbital radius $R$, the inclination $i$, the position angle $\Omega$, and the enclosed mass $M_{enc}$.

%------------------------------
\subsection{Modeling the polarization loop}
\label{subsec:analysis_modeling_the_polarization_loop}
%------------------------------

The polarization signal of an orbiting hot spot probes the magnetic field around Sgr~A* \citep{Broderick_2005,GRAVITY_2018b,GRAVITY_2020c,GRAVITY_2020d,Wielgus_2022,Michail_2023}. 
Simulations and analytical methods have been used by \cite{GRAVITY_2020c} and \cite{Narayan_2021} to investigate the polarization changes associated with such hotspots. A dominantly toroidal or radial field lead in general to two full loops in the $Q-U$ plane, while a dominantly poloidal field produces either no loops for edge-on geometries ($i \sim 90$), one loop for moderate inclinations ($i \sim \ang{30}/\ang{150}$), or two loops for face-on inclinations ($i \sim \ang{0}/\ang{180}$).
Hence, observing one loop per astrometric revolution severely constrains the configuration space and argues for a vertical magnetic field and moderate inclination. The repeated observations of single loops in the Q-U plane also indicate that the global magnetic field configuration has remained stable over recent years, consistent with the small changes in the circularly polarized emission at sub-millimeter wavelengths \citep{Munoz_2012}.

To derive the inclination constraint, we used the model introduced by \cite{Gelles_2021}. This semi-analytic model implements an equatorial point-like hot spot orbiting on a Keplerian orbit around a Kerr black hole with a local magnetic field. The model uses the semi-analytic solution for light rays \citep{Gralla_2020} to obtain a polarized image of the synchrotron emitting source. 

Since the model describes purely linear polarized emission, we extend it to take into account the observed polarization fraction in the data. Although coming from a point-source emitter, the model captures the fundamental features of extended emission regions \citep{GRAVITY_2020c,Vos_2022}. A more detailed description of the model can be found in Appendix~\ref{appendix:depolarized-hot-spot}.
	
The local magnetic field is given in cylindrical coordinates and allows a clear distinction between radial, toroidal, and poloidal magnetic field configurations. Other free parameters are the inclination of the orbit, the orbital radius, the local boost, and the spin of the black hole. The last only has a minor impact on the polarization signal \citep{Gelles_2021}, and hence we choose to work with zero-spin models. Further, we fix the radius to $9 R_g$ as this matches the observed astrometric and polarimetric periods. Using radii in the $8-10 R_g$ range does not alter our results (see Fig.~\ref{fig:gelles_model_mural}). Fixing the radius also fixes the Lorentz boosting factor to $\beta = 0.37$. The model does not take radiative transfer effects into account. Further, time delays and luminosity changes due to the combination of a non-flat spectral energy distribution and Doppler boosting are neglected.

\begin{figure}
	\centering
	\includegraphics[width=0.94\linewidth]{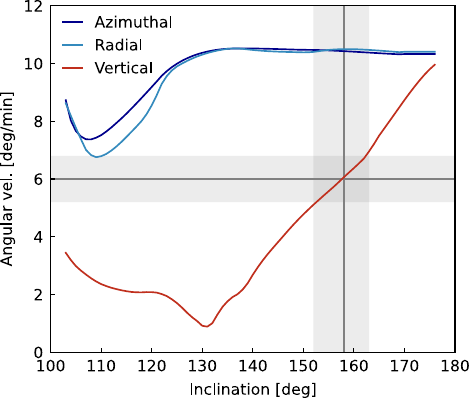}
	\caption{Comparison of the average angular velocity of the $Q-U$ loop for different magnetic field configurations as a function of inclination. Our measured values are consistent with a poloidal magnetic field and an inclination of $i = \SI{157}{\degree} \pm \SI{5}{\degree}$.}
	\label{fig:morphology_constraints}%
\end{figure}

As expected from the discussion above, the rotation speed distinguishes very well between the poloidal and toroidal/radial fields. Observing one polarimetric loop per astrometric loop only occurs for poloidal fields. The inclination can be constrained from the observed angular velocity with which the $Q-U$ loop is traversed (Fig.~\ref{fig:morphology_constraints}). The observed angular velocity of $\SI{6}{\degree/min}$ for the astrometry is only compatible with poloidal field configurations with an inclination of  $i = \SI{157}{\degree} \pm \SI{5}{\degree}$. 

\begin{figure}
	\centering
	\includegraphics[width=0.95\linewidth]{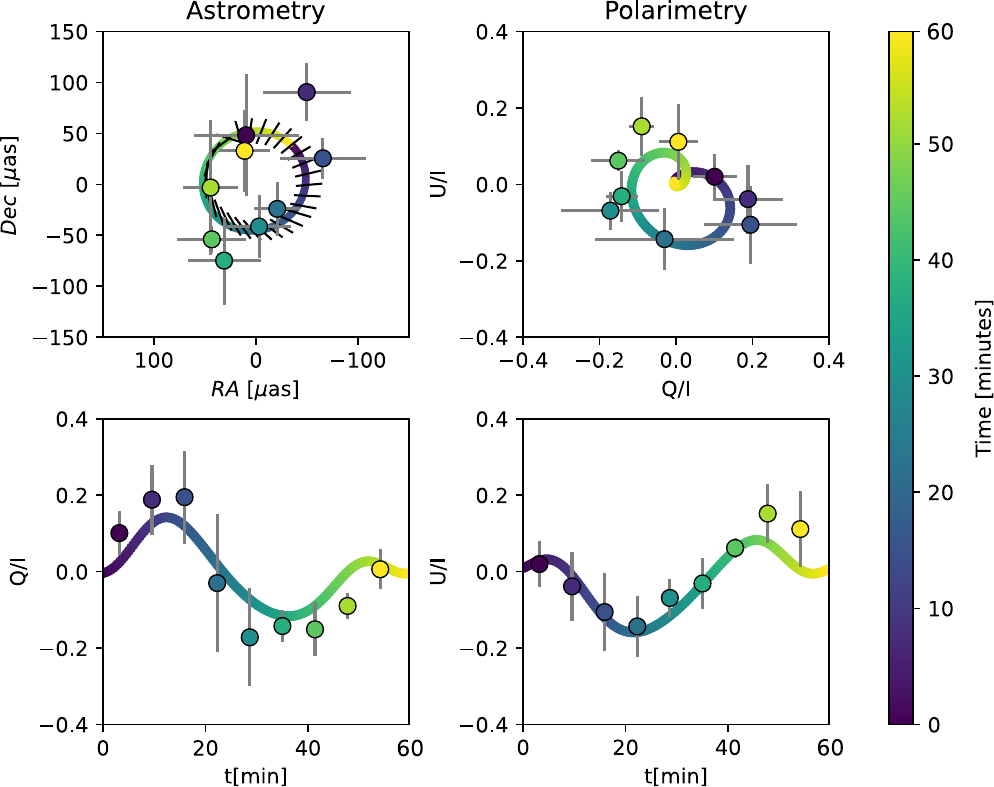}
	\caption{ Comparison of the polarization model with $R = 9 R_g$ and $i = \SI{157}{\degree}$ with the data from Fig.~\ref{fig:averaged_flares}.
	}
	\label{fig:polarization_fit}%
\end{figure}

Figure~\ref{fig:polarization_fit} shows that the model for $i = \SI{157}{\degree}$ is an excellent description of the observed polarization changes. This inclination matches the one measured by \cite{GRAVITY_2018b} and the preferred inclination of the EHT ring \citep{EHT_2022}.
We allowed the position angle on the sky and the arbitrary zero point in time to be adjusted, and picked the combination that yields the smallest $\chi^2$ to the data (Fig.~\ref{fig:chi_squared_fit}).
The best-matching position angle is $PA = \SI{25}{\degree}$.
This value is consistent with the results from \cite{GRAVITY_2020c}, \cite{Wielgus_2022}, and \cite{Michail_2023}.

In summary, the analysis of the polarization loops gives robust constraints and a result compatible with the astrometric signature. In the following, we combine this information in a combined fit.

%------------------------------
\subsection{Combined fit}
\label{subsection:analysis_combined_fit}
%------------------------------
%
\begin{figure*}
	\centering
	\includegraphics[width=0.9\linewidth]{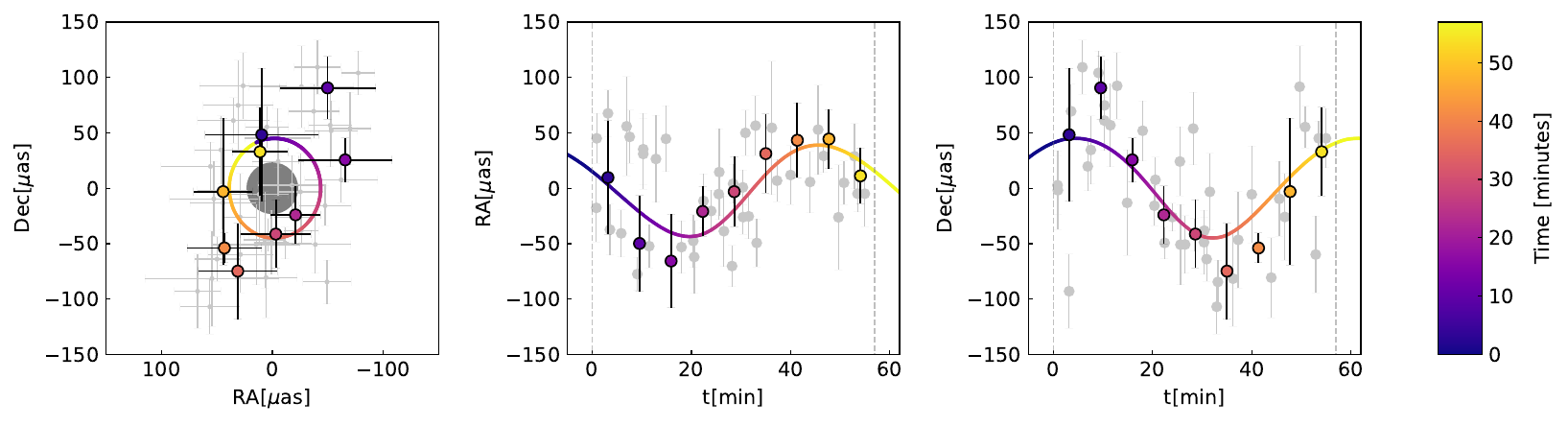}
	\caption{Combined fit of the astrometric flare data, taking into account the polarimetry constraints. Left: On-sky motion. The gray disk corresponds to the shadow size of a Schwarzschild black hole $3 \sqrt{3}R_g$. Middle and right panels: Individual coordinates as a function of time. The gray data points are the full data set and the colored points are bins of five minutes.}
	\label{fig:astrometry_fit}%
\end{figure*}
\begin{figure*}
	%\sidecaption
	\centering
	\includegraphics[width=0.8\linewidth]{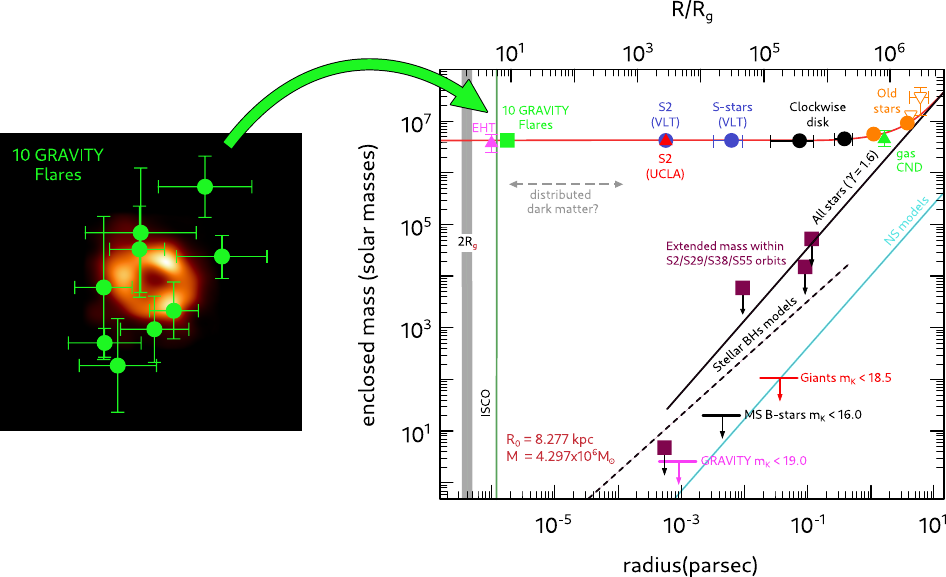}
	\caption{Central mass distribution in the GC. Right: Enclosed mass in the GC as a function of radius. The flare motions constrain the black hole mass of $M = (4.2 \pm 2.0) \times \SI{e6}{\solarmass}$ to be enclosed within less than 9 $R_g$. Out to \SI{e6}{} $R_g$ the potential is dominated by the MBH, as the comparison with theoretical estimates shows (solid and dashed lines). Left: Astrometric flare data overlayed on the EHT observation of Sgr~A* \citep{EHT_2022} that provides a constraint at a similar radius.%\vspace{3.4cm}
		}
	\label{fig:genzel_plot}%
\end{figure*}
\begin{figure*}
	\centering
	\includegraphics[width=0.8\linewidth]{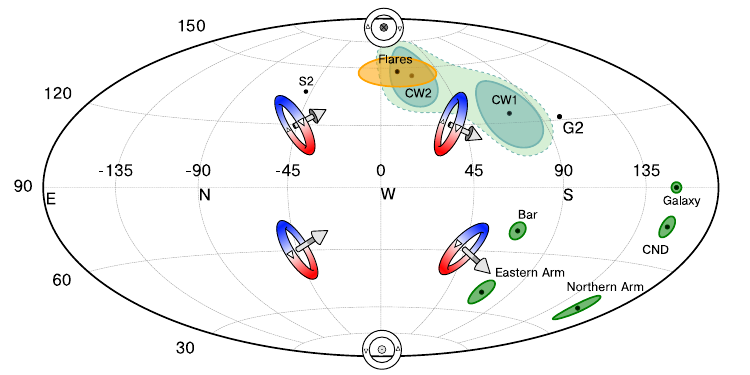}
	\caption{Compilation of orientations of angular momentum vectors of different dynamical structures in the GC. The colored tori illustrate the definitions of the angles. Inclinations between 0° and 90° correspond to counterclockwise motion on the sky (bottom half) and 90° - 180° to clockwise motion (upper half). The vectors of the clockwise stellar disk (shown are the features called CW1 and CW2 in \citealt{von_Fellenberg_2022}; see also \citealt{Paumard_2006}, \citealt{Lu_2008}), of the Sgr~A* flares (this work), and of the gas cloud G2 \citep{Gillessen_2011} are very close to each other. Other features shown are the rotation of the Galaxy, as well as the Circumnuclear Disk (CND) in the GC and the Eastern and Northern Arm gas features. The orientation of the flare angular momentum vector is consistent with the clockwise disk features.}
	\label{fig:orientation_plot}%
\end{figure*}

Our explicit combined fit (Fig.~\ref{fig:astrometry_fit} and Fig.~\ref{fig:corner_plot}) uses the astrometric data and the constraints derived from the polarimetry. We obtained the set of parameters presented in Table \ref{tab:best_fit_values}. 
\begin{table}
	\caption[]{Posterior mean values of the astrometry fit. The model corresponding to this set of parameters is displayed in Fig.~\ref{fig:astrometry_fit} (for details of the Bayesian analysis see Appendix \ref{appendix:fit}).}
	\label{tab:best_fit_values}
	\centering
	\begin{tabular}{cc}
	\hline \hline
	\noalign{\smallskip}
	Parameter & Posterior mean \\
	\noalign{\smallskip}
	\hline
	\noalign{\smallskip}
	Radius & $R = 8.9^{+1.5}_{-1.3} R_g$ \\[3pt]
	Inclination & $i = 154.9^{ +\SI{4.6}{\degree} }_{ \SI{-4.6}{\degree}}$  \\[3pt]
	Position Angle          & $\Omega = 177.3^{ +\SI{24}{\degree} }_{ \SI{-23}{\degree}}$ \\[3pt]
	Enclosed Mass          & $M_{enc} = 4.2^{+1.2}_{-0.9} \times 10^6\solarmass$ \\ 
	\noalign{\smallskip}
	\hline
\end{tabular}
\end{table}
Within the uncertainties, the flare radius and the orientation agree with what \cite{GRAVITY_2018,GRAVITY_2020c,GRAVITY_2020d} found from the previous smaller data set. The novel result is the mass constraint, that stems from the new data, in particular from the strong independent period constraint from the polarimetry. The value for the mass inside $9 R_g$ matches the simple estimate, and again also the mass derived using stellar orbits.

%============================================================

\section{Results}
\label{sec:results}

The most important result we find is that the mass of Sgr~A* measured from stellar orbits, $M = (4.297 \pm 0.012) \times \SI{e6}{\solarmass}$, is enclosed within 9 $R_g$, corresponding to 0.38 AU, or roughly the orbital radius of planet Mercury. A similarly strong constraint is obtained from the radio-VLBI image of Sgr~A* \citep{EHT_2022}. 

The stellar orbits with the sub-percent precise mass measurement, on the other hand, limit the size of Sgr~A* to be smaller than the smallest observed pericenter passages, which are those of S29 with 100 AU in 2021 (directly observed by \citealt{GRAVITY_2022}) and S14 with 26 AU in 2000 (well-determined orbit, but poorer coverage; \citealt{Gillessen_2017} and newer data). We show an updated version of the enclosed mass versus radius in Fig.~\ref{fig:genzel_plot}.

Other than the mass determination from the stellar orbits, both inner constraints are not assumption-free. The NIR flare constraint assumes that the motion of the emission is near Keplerian (i.e., governed by the forces of gravity), which is well motivated by 
%some 
accretion flow models \citep{Yuan_2003,Narayan_2023}. The EHT image needs to assume a certain (again well-motivated) structure of the accretion flow, as other gas configurations could mimic the ring-like appearance. Taken together, these two independent estimates make the case for a black hole ever more convincing.
Excluded are, for example, self-gravitating fermionic dark matter models with typical core radii of $\approx \SI{e-3}{pc} \approx \SI{200}{AU}$~\citep{Arguelles_2019,Becerra_Vergara_2020}.

The fact that the azimuthal speed of the hot spots is near Keplerian is, however, non-trivial. In low-density advection-dominated accretion flows (ADAFs), the gas motions generally are assumed to be sub-Keplerian \citep[e.g., ][]{Yuan_2014}. The same is not true for magnetically dominated flows, where the speed can depend on the spin of the black hole, and both sub- and super-Keplerian motions occur. We note that a number of non-Keplerian models have also been proposed. \cite{Aimar_2023} consider the outward motion of a hot spot along a conical trajectory, and \cite{Lin_2023} propose that, similar to coronal mass ejections on the Sun, Sgr~A* flux ropes are ejected out and filled with energetic electrons. \cite{Matsumoto_2020} discuss whether the observed motions are actually pattern motions.

In addition to the enclosed mass, we also measured the orientation of the flare orbits (see Fig.~\ref{fig:orientation_plot}). The measured orientation, which is the direction of orbital angular momentum, is close to that of the clockwise stellar disk \citep[][Fig.~9]{Levin_2003,Genzel_2003b,Paumard_2006,Lu_2008,von_Fellenberg_2022}, suggesting a physical connection. 

The stars that make up the disk are predominantly O--type and Wolf--Rayet--type stars that have strong winds. Simulations by \cite{Ressler_2020} indeed show that the almost spherical accretion flow, fueled by the stellar winds from disk stars, carries the initial angular momentum down to event horizon scales, where the gas settles into a disk-like structure. This disk can tilt by a moderate angle with respect to the initial angular momentum, perhaps responsible for the misalignment between the stellar disk and flare angular momentum directions in Fig.~\ref{fig:orientation_plot}.
The simulations also predict that the infalling gas settles into a magnetically arrested disk (MAD), which naturally carries a poloidal field geometry. Only for strong fields can the geometry withstand the dragging with the fluid that would lead to a toroidal field geometry.
Our data, favoring a poloidal field, directly support the MAD state. 
The MAD scenario is further supported by the estimated magnetic field strength (\SI{30}{}-\SI{80}{G}; \citealt{Ripperda_2020}) and measured density profile (see Fig.~6 in \citealt{Gillessen_2019} and references therein). \cite{McKinney_2013} and \cite{Ressler_2023} find that for MAD situations, the innermost disk-like structure tends to align with the spin axis of the black hole. The dynamics of NIR flares may, in that case, actually carry information on the black hole spin rather than just on the initial conditions of the inflow.

%============================================================

\begin{acknowledgements}
    We are very grateful to our funding agencies (MPG, ERC, CNRS [PNCG, PNGRAM], DFG, BMBF, Paris Observatory [CS, PhyFOG], Observatoire des Sciences de l'Univers de Grenoble, and the Fundação para a Ciência e Tecnologia), to ESO and the Paranal staff, and to the many scientific and technical staff members in our institutions, who helped to make NACO, SINFONI, and GRAVITY a reality. JS is supported by the Deutsche Forschungsgemeinschaft (DFG, German Research Foundation) under Germany's Excellence Strategy -- EXC- 2094 -- 390783311. A.A., A.F., P.G., and V.C. were supported by Fundação para a Ciência e a Tecnologia, with grants reference SFRH/BSAB/142940/2018, UIDB/00099/2020 and PTDC/FIS-AST/7002/2020. F.W. has received funding from the European Union's Horizon 2020 research and innovation programme under grant agreement No 101004719. Based on observations collected at the European Southern Observatory under the ESO programme IDs 109.22ZA.005, 109.22ZA.002, 105.20B2.004, 0103.B-0032(C), 0101.B-0576(E), 0101.B-0576(C).
    \end{acknowledgements}
    
\bibliography{./bibliography_short_journals}
%============================================================
%============================================================
\begin{appendix}
	
%------------------------------
\section{Rough mass estimate}
\label{appendix:mass_estimate}
%------------------------------

The period $P$ for an emitter on a circular orbit with radius $r_e$ is simply given by the common Kepler formula
\begin{equation}
	P^2 = \frac{4 \pi^2 r_e^3 }{GM} = \frac{4 \pi^2}{c^2} \frac{r_e^3}{R_g} \; .
\end{equation}
Following \cite{Gralla_2019} and \cite{Gates_2020}, the observed loop radius on the sky $r_o$ is (to very good approximation) $r_o = r_e +R_g$, meaning that the loop appears enlarged by roughly $1 R_g$ due to the gravitational lensing. This yields
\begin{equation}
	R_g = \left(\frac{2 \pi}{P c}\right)^2 (r_o - R_g)^3 \; ,
\end{equation}
which, for an observed period and radius, can be solved for $R_g$, and thus yields a mass $M$.

%------------------------------
\section{Depolarized hot spot model}
\label{appendix:depolarized-hot-spot}	
%------------------------------

The model used in Sect.~\ref{sec:analysis} builds on the model presented in \cite{Gelles_2021}. 
The original model describes the non-fractional Stokes parameters $Q$ and $U$ for completely polarized light such that the total intensity of the hot spot is given by $\sqrt{Q^2 +U^2}$. 
Experimentally we measure both polarized and non-polarized intensities such that the total intensity is $I = I_p + I_{np}$. While we assume that the flux is dominated by the hot spot, it still is only $10-40$ $\%$ polarized~\citep{GRAVITY_2020c}. If the non-polarized intensity comes from some stochastic process (e.g., turbulence in the flow, chaotic local magnetic fields), the averaging process of the non-polarized intensity will average to a constant value $\langle I_{np}\rangle$ if enough flares are considered. The average fractional polarization will then be
\begin{equation}
	\frac{Q}{I} = \frac{Q}{I_p + \langle I_{np}\rangle} \quad , \quad \frac{U}{I} = \frac{U}{I_p + \langle I_{np}\rangle} \; .
\end{equation}
Although a simplification, this approach captures the fundamental features of the complex dynamics seen in more realistic hot spot models \citep[see][]{Vos_2022,Dexter_2020}. 

\begin{figure}[]
	\centering
	\includegraphics[width=0.98\linewidth]{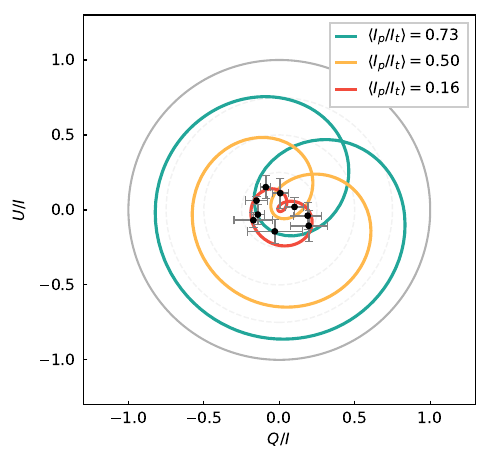}
	\caption{Effect of adding a constant depolarization term to the total intensity predicted by the hot spot model. The data points represent the averaged polarization data shown in Fig.~\ref{fig:averaged_flares}, and the red line the model whose average polarization matches that of the observed value. The inclination and position angle used are respectively $ i = \SI{157}{\degree} $ and $PA = \SI{25}{\degree}$.}
	\label{fig:fractional_polarization_example}%
\end{figure}

An example of how adding the non-polarized emission component can be used to match the observational average fractional polarization 
%\
\begin{equation}
	\left \langle \frac{I_p}{I} \right \rangle= 
	\left \langle \frac{\sqrt{Q^2 + U^2}}{I} \right \rangle
\end{equation}
is shown in Fig.~\ref{fig:fractional_polarization_example}. Matching the average polarization fraction of the data naturally represents the data points.

%\FloatBarrier

%%%%%%%%%%%%%%%%%%%%%%%%%%%%%%%%%%%%%%%%%%%
\section{QU loop morphology}
\label{appendix:QU_loop_morphology}
%%%%%%%%%%%%%%%%%%%%%%%%%%%%%%%%%%%%%%%%%%%

\begin{figure}[]
	\centering
	\includegraphics[width=\linewidth]{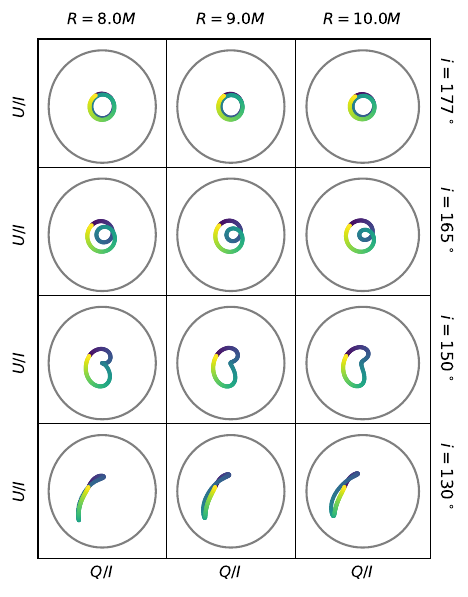}
	\caption{Illustration of Q-U loops for different radii and inclinations for the case of a poloidal field geometry. The color-coding indicates time, as in Fig.~\ref{fig:averaged_flares}. A single loop occurs for a narrow range of almost face-on inclinations. The gray circle represents unitary fractional polarization, and the average depolarized emission is $\langle I_{np} \rangle = 0.3$ for all plots.}
	\label{fig:gelles_model_mural}%
\end{figure}

\begin{figure}[]
	\centering
	\includegraphics[width=0.99\linewidth]{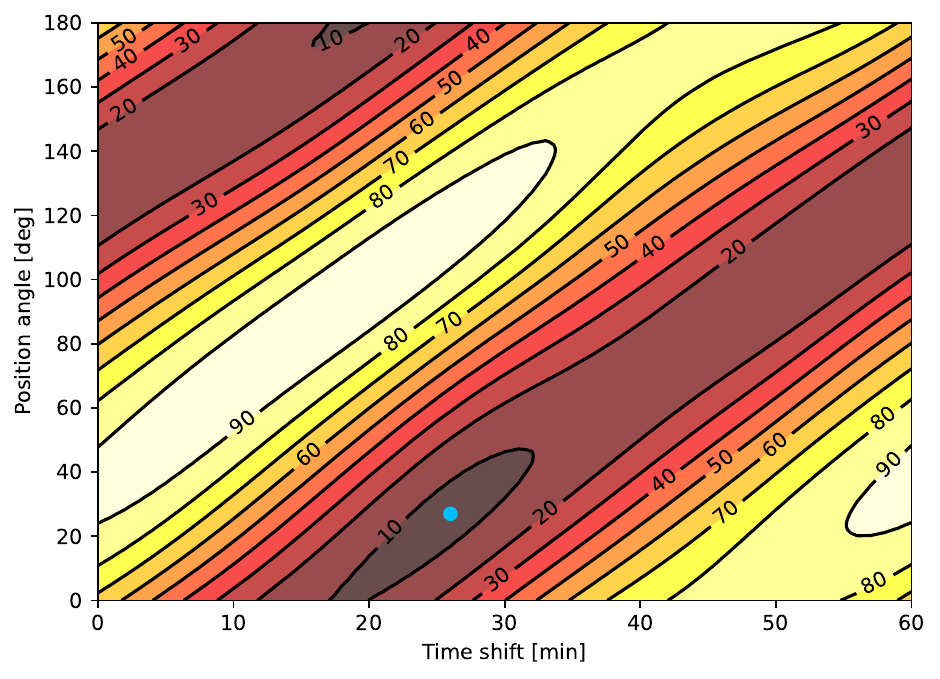}
	\caption{$\chi^2$ contour map for the polarization model as a function of position angle and time shift for a fixed radius of $9R_g$. The blue dot represents the minimum at (27 min, \SI{25}{\degree}). A slight degeneracy between the two variables exists since, for each position angle, there is a corresponding optimum time shift that adjusts the model to start at the right time. The corresponding model is displayed in Fig.~\ref{fig:polarization_fit}. }
	\label{fig:chi_squared_fit}%
\end{figure}

Regardless of the depolarization fraction used in the depolarized hot spot model (see Appendix \ref{appendix:depolarized-hot-spot}), the predicted electric field position angle (EVPA) is always given by
\begin{equation}
	\text{EVPA} = \frac{1}{2} \arctan \left( \frac{U}{Q}   \right) \; ,
\end{equation}
and depends solely on the non-fractional Stokes parameters. The angular velocity with which the $Q-U$ loops are traversed (see Fig.~\ref{fig:gelles_model_mural}) is thus independent of the non-polarized emission, and makes it a robust value for the inclination constraint presented in Fig.~\ref{fig:morphology_constraints}. By taking the inclination that matches the $\SI{6}{\degree/min}$ and fixing the polarization emission such that the average fractional polarization matches the observed value, we can fit for the position angle and time starting point of the loop. The $\chi^2$ fit is displayed in Fig.~\ref{fig:chi_squared_fit} and the corresponding model in Fig.~\ref{fig:polarization_fit}.

%
%%%%%%%%%%%%%%%%%%%%%%%%%%%%%%%%%%%%%%%%%%%		
\section{Fitting model}
\label{appendix:fit}
\begin{figure}[h!]
	\centering
	\includegraphics[width=0.99\linewidth]{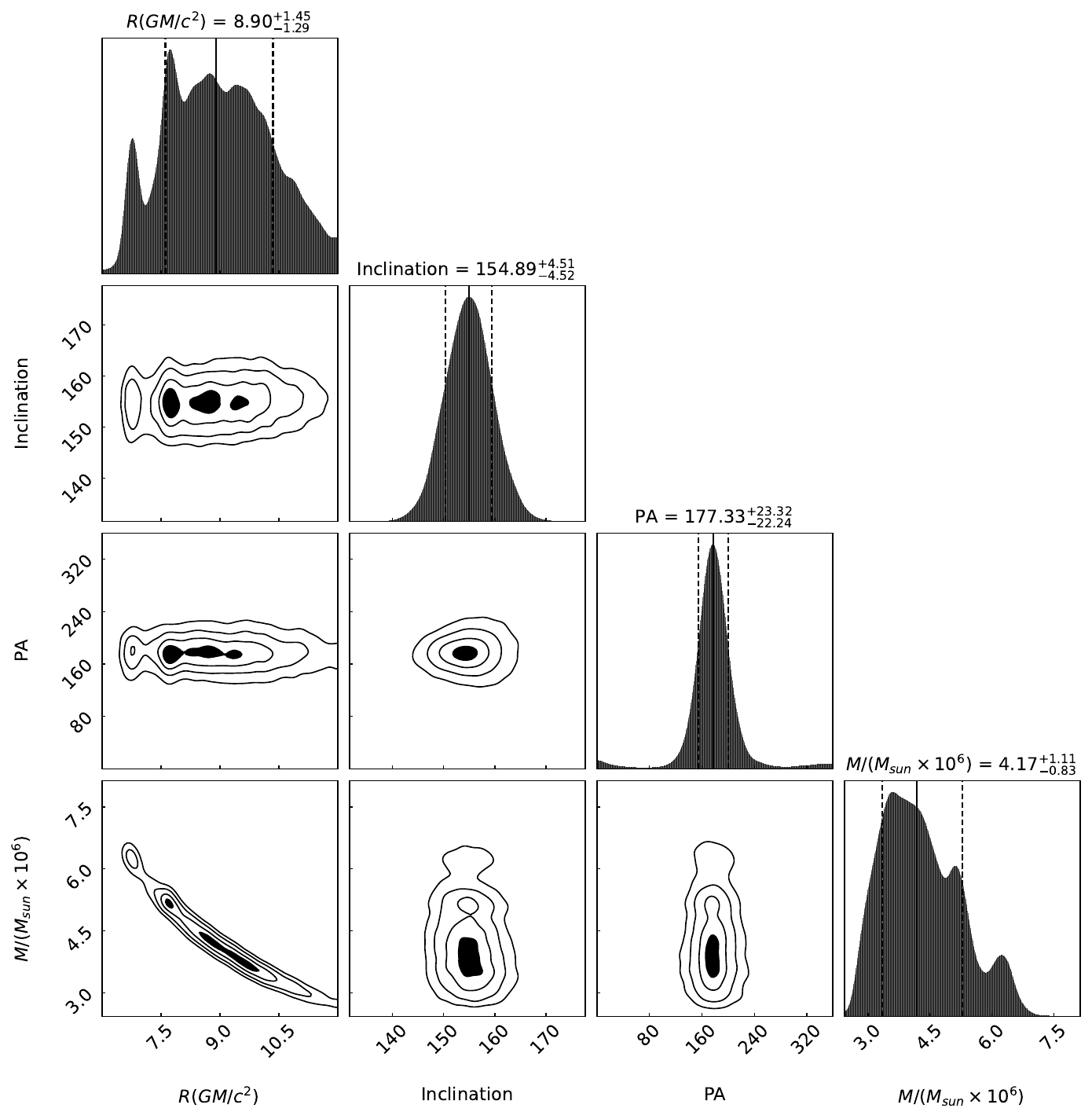}
	\caption{Two-dimensional posterior distributions of our combined fit with four free parameters: flare orbit radius, inclination, position angle, and enclosed mass. The posterior distribution of the radius parameter has multiple peaks, and hence the mean of the distribution is reported. There is a  correlation between mass and radius, due to the strong period constraint in our data.}
	\label{fig:corner_plot}%
\end{figure}

The astrometric fit presented in Sect. \ref{subsection:analysis_combined_fit} is similar to that presented in~\cite{GRAVITY_2020d}. In addition to the new constraints from the polarimetric data, we replaced the previously used $\chi^2$ fit with a nested sampling fit. The posterior distributions are shown in Fig.~\ref{fig:corner_plot} and the corresponding model in Fig.~\ref{fig:astrometry_fit}. We used flat priors for all quantities except the inclination, for which we used a sine prior. For the inclination and position angle we consider cyclic priors.

\end{appendix}

\end{document}